# The Effects of Vehicle-to-Infrastructure Communication Reliability on Performance of Signalized Intersection Traffic Control

Ilya Finkelberg, Tibor Petrov, Ayelet Gal-Tzur, Nina Zarkhin, Peter Počta, Tatiana Kováčiková, Ľuboš Buzna, Milan Dado, Senior member, IEEE, Tomer Toledo

*Abstract*—A vehicle-to-infrastructure communication can inform an intersection controller about the location and speed of connected vehicles. Recently, the design of adaptive intersection control algorithms that take advantage of this information evoked a lot of research, typically assuming a perfect communication. In this study, we explore possible effects of a temporal decrease in the reliability of the communication channel, on a throughput of a signalized intersection. We model road traffic and DSRC-VANET communication by integrating the well-established traffic and communication simulation tools (Vissim and OMNeT++). Comparisons of the perfect communication conditions with challenging, but realistic conditions involving communication distortions show that the adaptive intersection control may suffer from significantly increased average delays of vehicles. The level of delays is largely independent of whether the communication distortions affect all or only a single intersection approach. For some signal groups, the average vehicle delays are significantly increased, while for the others they are decreased, leading to the disbalance and unfairness. We show that the data received in the previous time intervals and simple assumptions about the vehicle movements can be used to estimate the lost data. The compensation of the communication distortions affecting a single intersection approach decreases the average vehicle delays. When the communication distortions impacting all the intersection approaches are compensated for, the vehicle delays are even set back to the levels comparable with the perfect communication conditions. Overall, the results demonstrate that the impact of the communication distortions should be considered in the design of the adaptive intersection control algorithms.

*Index Terms*—Communication distortions, Reconstruction of incomplete communication, Signalized intersection control, Vehicle delays, Vehicle-to-infrastructure connectivity.

## I. Introduction

As the population grows, traffic congestion in urban centers increases, leading to longer delays, increased emissions and decreased traffic safety. The introduction of connected vehicle (CV) technology may provide near future solutions to improve performance through driving assistance, collision avoidance and traffic management applications [1]. CV telematics is a rich source of real-time information about the state of the traffic system that may be used to enhance the efficiency of traffic management and control, most notably intersection signal control.

In recent years, various traffic signal control algorithms utilizing the CV data have been developed [2]. Ref. [3] proposed a distributed coordinated signal control methodology, where the signal control parameters are optimized to reduce vehicle delay and increase throughput. Ref. [4] developed a two-level optimization model where the sequence of signal stages and their durations are optimized based on the incoming CV data. Another recent work [5] proposes a heuristic approach to determine optimal platoon discharge order at a signalized intersection to minimize vehicle delay.

A common assumption in these studies is that the communication, both Vehicle-to-Vehicle (V2V) and Vehicle-to-Infrastructure (V2I), is very reliable, i.e. without any loss or delay. This assumption is rarely realistic. It is worth noting here that a communication performance is affected by a variety of factors, such as the separation distance, vehicle speed, number of vehicles within the communication range, physical obstacles (e.g. buildings, trees), presence of interference sources and weather conditions [6].

Currently, there are two dominant branches of communication technologies in the context of connected vehicles: Dedicated Short-Range Communication-based Vehicular Ad hoc Network (DSRC-VANET) technologies that use Wi-Fi-like IEEE 802.11p communication standard and

This work has been submitted to the IEEE for possible publication. Copyright may be transferred without notice, after which this version may no longer be accessible. Manuscript received on ………….. This research was supported by a grant from the Ministry of Science & Technology, Israel and by the Slovak Research and Development Agency under the contract no. SK-IL-RD-18-005. L.B. was partly supported by the research grants: VEGA 1/0089/19 Data analysis methods and decisions support tools for service systems supporting electric vehicles, and APVV-19-0441 Allocation of limited resources to public service systems with conflicting quality criteria. The authors would like to thank Dr. Marek Drličiak and Prof. Ján Čelko for their invaluable support in facilitating the use of simulation tools essential for this research.

I. Finkelberg, A. Gal-Tzur, N. Zarkhin and T. Toledo are with the Transportation Research Institute, Technion – Israel Institute of Technology, IL-32000 Haifa, Israel (e-mail: ilya.f@technion.ac.il; galtzur@technion.ac.il; ninaz@technion.ac.il; toledo@technion.ac.il).

T. Petrov is with the Department of International Research Projects - ERAdiate+, University of Žilina, Univerzitná 8215/1, SK-010 26 Žilina, Slovakia (e-mail: tibor.petrov@erachair.uniza.sk).

T. Kováčiková is with the the Department of International Research Projects - ERAdiate+ and the Department of Information Networks, University of Žilina, Univerzitná 8215/1, SK-010 26 Žilina, Slovakia (e-mail: tatiana.kovacikova@erachair.uniza.sk).

P. Počta, M. Dado are with the Department of Multimedia and Information-Communication Technologies, University of Žilina, Univerzitná 8215/1, SK-010 26 Žilina, Slovakia (e-mail: peter.pocta@feit.uniza.sk; milan.dado@feit.uniza.sk).

Ľ. Buzna is with the Department of Mathematical Methods and Operations Research and the Department of International Research Projects - ERAdiate+, University of Žilina, Univerzitná 8215/1, SK-010 26 Žilina, Slovakia (e-mail: lubos.buzna@fri.uniza.sk).

Cellular Vehicle-to-Everything (C-V2X) technologies that are built on mobile cellular networks, such as LTE or 5G in the future. A number of studies have evaluated the performance of the communication technologies. Ref. [7] showed that LTE-based C-V2X can achieve the same block error ratio with lower Signal-to-Noise Ratio (SNR) than the DSRC-VANET systems on a larger coverage area. Similar results are presented in [8] for message exchange within a platoon of trucks on a highway. The authors conclude that LTE C-V2X outperforms the DSRC-VANET in terms of communication reliability while the DSRC-VANET technology has an advantage in a communication latency. It is worth noting here that LTE C-V2X was first standardized in 2017. So, [9] points out that it is still not market ready. DSRC-VANET is more mature and already popular. It is therefore used throughout this research.

Regardless of the technology used, the communication reliability of connected vehicles is far from ideal. Ref. [10] showed that when considering a realistic signal propagation model and a moderate distance of 300 meters between a transmitter and a receiver, the application-level reliability of the DSRC-based communication is not more than 55% for delay sensitive safety applications. Similarly, [11] concluded that the placement of the Onboard Unit (OBU) antenna inside a vehicle or on its roof, as well as terrain topology play a major role in communication reliability. Even with the OBU antenna installed on the vehicle's roof, the message delivery ratio ranges from 71 to 88% in an urban environment. Furthermore, the authors conclude that even on a straight road, altitude variations between the communicating vehicles can significantly reduce the reliability of V2V communication.

Real-world applications of connected vehicles are still rare. Therefore, their evaluations are mostly carried out using simulation models. Ref. [12] calibrated communication simulation models with empirical data obtained from the CV testbed operated by the Federal Highway Administration. Refs. [13], [14], [15], [10] used a hybrid simulation model that combines a traffic micro-simulation model and a communication network simulator. These studies focus on the performance of the communications network and do not consider the effects of imperfect data transmission on traffic performance. Few studies considered the bi-directional dependency between the two systems. Ref. [16] presented a communication fault tolerant algorithm for control of a platoon of autonomous CVs. Ref. [17] study the impact of security attacks on traffic control performance using integrated communications and traffic simulation.

In summary, the literature shows that traffic control and traffic flow conditions both affect and are affected by the CV communication network reliability. However, the dependence between the two systems is often ignored in their performance evaluations, which commonly assume perfect communications. Therefore, the necessary modules to detect and compensate for communication faults are missing.

In this paper, an integrated evaluation framework that combines a traffic flow and control model with modeling of the communication network is presented. The developed simulation framework uses the VISSIM microscopic traffic simulation model and OMNET++ communication network simulator. The integrated framework is used to evaluate a control algorithm for a signalized intersection serving CVs and takes into account the impact of communication distortions on its performance. In the context of this paper communication distortions refer to a short-term disruption of message delivery caused by a challenging communication environment, i.e. decreased reliability of the communication channel. The long-term unavailability of the communication system caused by hardware failures is not considered. The results show a potential for substantial deterioration in the performance due to communication distortions. They also demonstrate the necessity and possible benefits of incorporating mechanisms to take into account communications distortions within the signal control logic.

The rest of this paper is organized as follows: the next section presents the integrated traffic and communications modelling framework and the main assumptions made in setting it up. Section III introduces the intersection signal control algorithm and a mechanism, which the algorithm incorporates, to take into account the possibility of communication distortions. Section IV describes the case study and design of simulation experiments. Section V presents the results of simulation experiments and their analysis. Finally, discussion and conclusion are presented in Section VI.

## II. INTEGRATED TRAFFIC AND COMMUNICATIONS MODEL

### A. Overall simulation framework

As noted above, simulation modeling is currently the only viable approach to evaluate the performance of traffic systems that include CVs and to consider the communications system. Fig. 1 shows the flowchart of the integrated traffic and communications simulation framework. It supports modeling of the real-time bi-directional data exchange between the traffic and communications models. After initialization of the simulation process, at the beginning of every time step, the traffic simulation model receives a list of communication messages from CVs (e.g. their identification numbers, locations, intended maneuver at the intersection and so on) that were received by the road side unit (RSU) in the previous time interval. Within the traffic simulation step, control actions (e.g. traffic light indications) are determined using this information and the vehicles in the model are advanced based on traffic conditions and the prevailing control states. At the end of the traffic simulation time step, a list of communication requests from CVs is generated. These requests are sent to the communications network simulation model. The model simulates the message exchange between the CVs and RSU within the same time step. It generates a list of communication messages that were successfully delivered to the RSU. The simulation clock is advanced and the communications list is used by the traffic simulation model in the next simulation time step. The simulation loop continues until the simulation end time is reached.

The implementation of the framework in this research used the VISSIM traffic simulation model. The traffic control algorithm described in Section III was implemented as an external computer program and integrated into VISSIM with VISCOM module (VISSIM Component Object Model). The Objective Modular Network Testbed in C++ (OMNeT++) was used to simulate the communication network. Communication protocols in OMNeT++ were simulated with the INET simulation framework [18]. The exchange of information between the traffic and communication simulators was based on shared files.

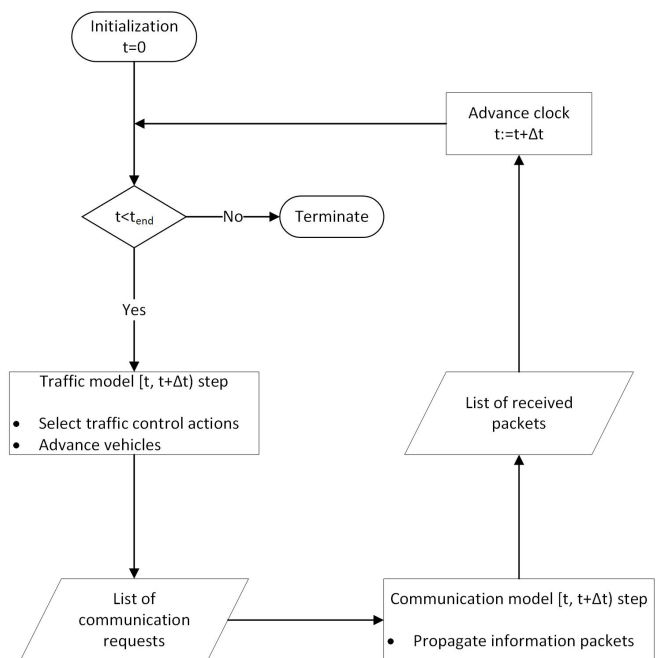

Fig. 1. Flowchart of the integrated traffic and communications simulation framework.

### B. Communication network modelling

The communication network model consists of three components: vehicles equipped with a communication module, communication channel and the RSU interconnected with the Intersection Controller (IC) as shown in Fig. 2. The following features are assumed for the communication network:

- Both the RSU and the vehicles use standardized IEEE 802.11p wireless interfaces operating in the 5.9 GHz frequency band as specified in ETSI EN 302 663 [19].
- The communication messages are transmitted using the IEEE 802.11p-based DSRC-VANET technology.
- All the vehicles and the RSU are mutually compatible and use standardized interfaces, communication protocols and messages.
- All network devices included in the model use idealized omnidirectional antennas, which neither amplify the transmitter output, nor introduce any loss to the system.
- The default DSRC-VANET data rate and underlying error protection mechanisms (e.g. modulation, channel coding) are used [19].
- All the vehicles are in a line-of-sight to an RSU antenna.
- Since the RSU and IC are expected to be connected via a wired link, the connection is assumed to be lossless, reliable and causing a negligible communication delay compared to the wireless links.

Each vehicle in the OMNeT++ simulations is represented by a compound module using IEEE 802.11 Physical (PHY) and Medium Access (MAC) layers. Each vehicle runs a User Datagram Protocol (UDP) application and uses standardized Cooperative Awareness Messages (CAMs) to transmit their telemetric data to IC via RSU. Each vehicle sends to RSU one CAM per second to update the IC about its current state. To avoid MAC layer collisions, a start time of the first message transmission is selected randomly at the time when the vehicle enters the simulation. All the following messages are regularly transmitted with a separation of one second. This CAM generation frequency reflects the requirements determined by preliminary simulation experiments and is in compliance with CAM generation frequencies specified in ETSI EN 302 637-2 [20]. Similarly, preliminary simulation experiments confirmed that it is sufficient if CAM transmissions occur in the close vicinity of the intersection, hence, the communication messages are sent directly from the vehicles to the RSU by one hop. Therefore, routing of messages from the sender to the receiver via intermediate vehicles is not considered. The wireless communication channel is an ETSI EN 302 663 [19] compliant Control Channel (CCH). The CCH is reserved for CAM broadcasts and safety-related V2X applications according to ETSI TS 102 724 [21].

Fig. 2 also shows the sources of communication distortions considered in the model. These sources may induce loss of the CAM message transmitted over the DSRC-VANET communication network. Besides natural attenuation, the quality of the communication can be affected by noise, interference and fading [22]. Noise is any unwanted fluctuation in a signal, which obstructs and masks the desired signal [6]. Its sources may be either natural or man-made. Interference signals are usually man-made and, unlike noise, their source can be easily identified. Possible interference sources may include frequency reuse, signals in adjacent channels with components outside their allocated frequency range or colliding transmissions coming from transmitters using the same channel [22]. Fading is a variable attenuation of the signal that may result from obstacle shadowing, atmospheric disturbances, or multiple propagation paths of the signal.

In the model, the level of overall noise signals (natural and man-made) present in the communication channel is captured by "background noise" power level, which is kept constant in all simulation experiments. The attenuation of the signal, i.e. the path loss, experienced during its propagation over the communication environment is modelled by the well-established Two-Ray Interference path loss model [23]. The level of signal distortion introduced by fading and interference is quantified by a Signal-to-Noise Ratio (SNR) penalty [24]. The SNR penalty describes a change of SNR of the transmitted signals at the receiving point (the RSU) compared to a Baseline situation, which in this case is when only the background noise and signal attenuation are considered. The combined effect of fading and interference is modelled by varying the sensitivity of the receivers located at the RSU.

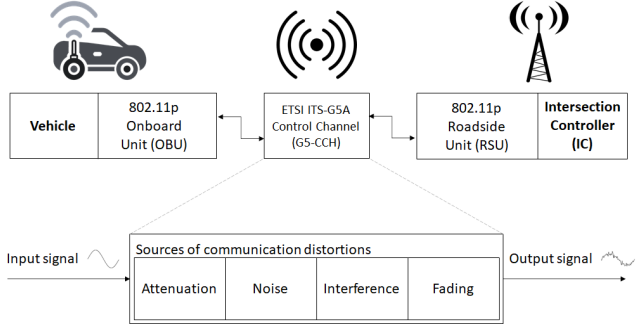

Fig. 2. Main components of the DSRC-VANET communication network and considered sources of communication distortions.

## III. TRAFFIC SIGNAL CONTROL SCHEME

The focus of the reported case study is on evaluation of possible impacts of communications distortions on the traffic signal control performance. Therefore, it is advantageous to use a transparent control scheme with a clear set of rules that are easy to diagnose. This archetypical model can also be easily adapted and extended to include realistic constraints and additional modules for improved performance and fault redundancy. The heuristic traffic control logic used in this research is an extension of the algorithm proposed by [25]. The main idea is to select control stages and their durations in real-time based on a weighted score reflecting vehicles' demand on the lanes associated with the stages. The weights used in this score are higher for vehicles that report shorter distances from the intersection in an attempt to prioritize them and increase the total throughput. Similar stage prioritizations are also proposed elsewhere, for example, in [26] and [27].

The basic principles and definitions of the traffic control logic are consistent with common practice. Vehicle movements in the intersection are organized together into a set $G$ of Signal Groups (SG). A subset of non-conflicting SGs that can receive green light at the same time constitutes a stage. For a more flexible control, a set of feasible stages $S$ is defined. A signal group can belong to one or more stages. The association of a signal groups $g \in G$ to a stage $s \in S$ is modeled by an incidence matrix $B$ with entries $b(g,s)$, which take value 1, if signal group $g$ belongs to stage $s$ and 0 otherwise. Both signal groups and stages are pre-defined in the design process and remain fixed during its operation.

When the signal control is operational, after each second, one of the following control decisions is reached – extend green to the current active stage or end the green for a current stage and switch to another one. This decision is based on the following conditions:

- The green period length of stage $s \in S$ is bounded by the minimum $l_s^{min}$ and maximum $l_s^{max}$ values. No stage switch may occur before the minimum duration is reached and a stage switch is mandatory when the maximum length is reached.
- When the duration of the green period for the active stage $s \in S$ is between $l_s^{min}$ and $l_s^{max}$, the green time for this stage is extended by one second, if the discharge flow is still high. Specifically, the green time is extended if the current headway $h_g$ at the stop line (gap time) is lower than a predefined constant $h^{max}$ for at least one lane belonging to an active SG $g \in G$. Otherwise, the current stage is terminated and a stage switch is initiated.
- On a switch to the next stage, in order to provide a safe transition, an all-red intergreen period with a predefined constant duration is activated.
- In order to limit the maximum possible delay, a control cycle is used. In each cycle, all non-zero demand signal groups must receive green light, and each stage can only be activated once.

When a current stage is terminated, the selection of the next stage to switch to is based on a weighted demand score derived from the information received from CVs. It is assumed that all vehicles are connected and that they constantly transmit their telemetric data to the intersection controller. This data includes their unique ID, location coordinates, speed, current lane and intended movement in the intersection. The intersection controller (IC) identifies the current set $C$ of vehicles that intend to cross the intersection. It assigns each CV $c \in C$, to the relevant SG. This assignment is captured by a matrix $A$ with entries $a(c,g)$, which take value 1 if vehicle $c$ is assigned to signal group $g$ and 0 otherwise. The IC than calculates the current distances $d_{ct}$ of the CVs from the stop line and assigns them scores based on these distances:

$$w_{ct} = max\left\{0, 1 - \frac{d_{ct}}{d_{max}}\right\}. \quad (1)$$

Where, $w_{ct}$ is the score of the vehicle $c$ at time $t$. The parameter $d_{max}$ is the maximum detection distance from the intersection. The scores $w_{gt}$ of the signal group $g \in G$ are calculated as the sum of scores of the associated vehicles:

$$w_{gt} = \sum_{c \in C} w_{ct} a(c,g). \quad (2)$$

Finally, the stage scores $w_{st}$ are calculated as the sum of scores of the signal groups that are active in the stage:

$$w_{st} = \sum_{g \in G} w_{gt} b(g,s). \quad (3)$$

When the current stage is terminated, the algorithm will initiate a transition to the stage with the currently highest weighted score, denoted as $w_s^{max}$, on a condition that this stage includes at least one SG that has not yet been served in the current cycle. Thus, in most cases the green light is granted to the stage with the largest number of vehicles close to the stop line, which aims to increase the efficiency of the green light utilization and yield higher throughputs.

In addition, a control adaptation mechanism is implemented. The maximum duration of a stage $s$, $l_s^{max}$, is updated in each cycle to better accommodate current demands. This approach is a core principle in classic adaptive traffic control algorithms, e.g. SCATS [28]. It allows congested stages with higher scores the potential for longer green durations in the next cycle.

In order to calculate the maximum green duration for each stage and for the purpose of limiting cycle length to a reasonable maximum, a constant $e$ representing the total available green extension is defined.

At the time when a decision is made to allocate the green time to stage $s$, the value $w_s^{max}$ is stored in the memory. At the end of the cycle, in order to divide available green times proportionally between the stages, $w_s^{max}$ is used to calculate the score $w_s^e$ for each stage $s \in S$ as:

$$w_s^e = \frac{w_s^{max}}{\Sigma_{s \in S} w_s^{max}}. \quad (4)$$

The maximum length of green period for stage $s$, used in the next cycle, is then calculated as:

$$l_s^{max} = l_s^{min} + w_s^e \, e. \quad (5)$$

As shown in the literature review, most of the research on CV-based traffic control assumes full and perfect information is available. Therefore, mechanisms to handle communication distortions are missing. The first step towards overcoming this problem is to identify communication distortions. In each time interval, the RSU received messages from CVs, which include their unique ID numbers. $C_t$ is the list of these vehicles. The weighted scores for these vehicles are calculated based on their reported location data as described above. $C_{t-1}$ is the list of the vehicles that communicated in the previous interval. Vehicles in $C_{t-1}$ but not in $C_t$ are suspected to have failed to communicate. The locations of these vehicles are extrapolated from their previous received reports. In order to simplify the extrapolation, it is assumed that they travel at a constant speed and do not change lanes, so that their position is constrained by that of their leader. Under these assumptions, the estimated current distance and speed of a CV are given by:

$$d_{ct} = max\{d_{c,t-1} - v_{c,t-1}, d_{c-1,t} + x_{min}\}, \quad (6)$$

$$v_{ct} = v_{c,t-1}. \quad (7)$$

Where, $v_{ct}$ and $v_{c,t-1}$ are the speeds of the vehicle in the current and previous time steps. $d_{c-1,t}$ is the position of the vehicle in front of vehicle $c$ on the same lane. $x_{min}$ is the minimum assumed distance between consecutive vehicles on the same lane.

Vehicles that are estimated to still be on the approach to the intersection ($d_{ct} >= 0$) are added the set $C_t$. $d_{ct} < 0$ implies that the vehicle crossed the intersection. If the light for its movement is currently green, the vehicle is not considered further. However, if the light is currently red, it is added the set $C_t$, and assumed to be at the stop line ($d_{ct} = 0$).

## IV. CASE STUDY

The impact of communication distortions on the performance of the signal control is demonstrated using a simulation case study of a generic four-legged intersection, as shown in Fig. 3a. The traffic movements on each approach to the intersection are organized in two signal groups: one for through and right turn (TR) traffic and the other for the left turn (L) traffic. These signal groups are assigned to a total of eight possible stages, shown in Fig. 3b.

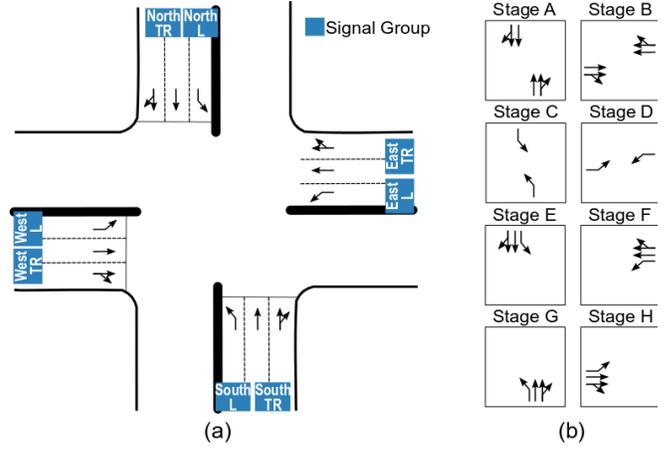

Fig. 3. Layout of the case study intersection and possible signal stages in the intersection.

TABLE I
VEHICLE FLOWS AT THE INTERSECTION

| Approach | Volume [veh/hr] | Signal Group | Volume [veh/hr] |
|---|---|---|---|
| North | 350 | North TR | 213 |
|  |  | North L | 137 |
| South | 800 | South TR | 640 |
|  |  | South L | 160 |
| East | 900 | East TR | 648 |
|  |  | East L | 252 |
| West | 850 | West TR | 748 |
|  |  | West | 102 |

TABLE II
TRAFFIC CONTROL PARAMETER VALUES

| Parameter | Description | Value |
|---|---|---|
| $l_s^{min}$ | Minimum green time | 6 s |
| $I$ | Interstage duration | 10 s |
| $e$ | Total Available green extension | 56 s |
| $h^{max}$ | Gap time | 3 s |
| $d_{max}$ | Maximum detection range | 300 m |

The traffic demand volumes used in the simulations are presented in Table I. The demand is unbalanced with lower volumes on the North-South direction, and especially the North approach. These values reflect a degree of saturation of about 0.65.

The values used for the traffic control parameters discussed in Section III are presented in Table II.

Parameter values used in the communications model and their sources are shown in Table III. Message length was set to 300 bytes to accommodate the payload data fields (position, driving lane, distance to the intersection, movement in the intersection), the message header and message authentication overhead. In general, one antenna might not be sufficient to enable a line-of-sight communication for all the intersection legs. Therefore, an RSU with four independent antennas, one for each intersection leg was assumed. The antennas are located above the stop line.

TABLE III
COMMUNICATION MODEL PARAMETER VALUES

| Simulation model parameter | Value |
|---|---|
| Message generation frequency | 1 s [20] |
| Message length | 300 Bytes |
| Carrier frequency | 5.9 GHz [19] |
| Communication channel bandwidth | 10 MHz |
| DSRC-VANET maximum data rate | 6 Mbps [19] |
| Permittivity of asphalt at microwave frequencies | 4.75 [29] |
| RSU antenna height above the road surface | 5.897 m |
| Vehicle antenna height above the road surface | 1.895 m [23] |
| Transmitter power | 20 dBm [30] |
| Background noise level | -86 dBm |

The simulation experiments focus on three factors that affect the performance of the communications and traffic systems:

1. Communication environment - Three types of environments are considered: (i) A Baseline condition assumes perfect communications with no failures. The traffic control algorithm works with the actual locations of all vehicles. This condition is used as a benchmark to the traffic impacts of imperfect communications. In the two other conditions the impact of the communications are considered. (ii) A homogenous communications environment condition assumes similar communication characteristics on all approaches to the intersection. This does not necessarily imply an equal information loss on every approach, as the dynamic spatio-temporal state of the communicating vehicles impacts communication reliability. (iii) A heterogeneous communication environment condition considers uneven communications characteristics. A higher level of channel degrading factors is assumed on the West approach to the intersection. This condition was designed to examine the impact of asymmetry in the quality of information about traffic flows on the performance of the traffic control.

2. The level of SNR penalty captures the combined effect of fading and interference in a challenging communication environment. Ref. [31] showed that the power offset of the received signal in the 5.9 GHz frequency band introduced by the fading, i.e. the SNR penalty, can be as high as 30 dB. In the simulation experiments this value is used as a worst-case. The SNR penalty level is varied in four steps: 0, 20, 25 and 30 dB. In the homogeneous communication environment, the same SNR penalty is applied to all the intersection approaches. In the heterogeneous communication environment, these SNR penalty values are applied only to the West approach. The other approaches face 0 dB SNR penalties. Thus, the homogeneous and heterogeneous communication environments are identical only when the SNR penalty is 0 dB.

3. Correction of communication distortions - Two variants of the control algorithm are considered: One that does not correct communication distortions, and the other that does, using the estimation approach described in Equations (6)-(7).

The combination of the values of these three factors define a total of 15 conditions. For each condition, 10 independent simulation replications were run. Each run includes a 10 minutes warm-up period used to populate the intersection with vehicles and a 30 minutes evaluation period.

## V. RESULTS

The performance of a communication network can be quantified by the message loss ratio (MLR), which is given by:

$$MLR = 1 - \frac{N^{rec}}{N^{snt}}, \quad (8)$$

where $N^{snt}$ and $N^{rec}$ are the numbers of sent and received messages, respectively.

Table IV shows the average MLR values over the simulation runs of the studied scenarios and their standard deviations. Note that homogenous and heterogenous scenarios with same SNR penalties are not directly comparable, as in the latter, the more challenging communication conditions were applied only to the West approach. MLR values are shown for the whole intersection, as well as for the West approach and the other approaches separately. The results show substantial loss of messages in all the cases. The loss is around 20%, with the 0 dB SNR penalties and reaches almost 80% at the 30 dB SNR penalty level. These MLR results are consistent with the ranges reported by [10] and [11]. The SNR penalties for different approaches are modelled independently, since it was assumed that each approach has a separate RSU antenna and interface. Thus, the MLR for the West approach is similar in the homogenous and the heterogenous scenarios, and the values for the other approaches in the heterogenous scenarios are similar to those of the 0 dB homogenous scenario. Finally, it is noted that the MLR for the scenarios with communication distortion corrections are consistently lower than those for the scenarios without correction. We attribute these differences to the changed vehicular flow dynamics. These differences are statistically significant at all the SNR penalty levels and are larger for the 30 dB condition.

The communication distortions translate into a degradation of the traffic system performance. Table V shows the average vehicle delays and their standard deviations in the various conditions. When the communication distortion correction is not applied, as the SNR penalty increases, initially only marginal and statistically insignificant increases in the vehicle delays over the Baseline condition are observed. Only with the highest, but still realistic, 30 dB SNR penalty, the increase in delays is large (22.0% and 20.7%, for the homogenous and heterogenous conditions, respectively) and statistically significant. This suggests the existence of a critical MLR value that when exceeded traffic performance is strongly impacted. At all the SNR penalty levels, the delays are similar for the homogeneous and heterogeneous disruption conditions.

These results also demonstrate the value of implementing a correction mechanism, such as the one presented in the previous

TABLE IV
MLR VALUES AND THEIR STANDARD DEVIATIONS IN THE VARIOUS CONDITIONS

| Condition | SNR penalty | MLR (%) and their standard deviations | | | | | |
|---|---|---|---|---|---|---|---|
| | | Without correction | | | With correction | | |
| | | All | West | Others | All | West | Others |
| Homogenous | 0 | 20.3 (0.25) | 23.4 (0.68) | 19.4 (0.32) | 18.2 (0.29) | 20.8 (0.90) | 17.2 (0.37) |
| | 20 | 36.9 (0.42) | 42.3 (0.76) | 35.4 (0.72) | 34.9 (0.49) | 39.6 (1.52) | 33.0 (0.49) |
| | 25 | 47.6 (0.46) | 55.3 (1.31) | 45.3 (0.58) | 46.3 (0.57) | 53.4 (1.20) | 43.4 (0.70) |
| | 30 | 64.7 (2.03) | 74.7 (2.53) | 61.4 (2.77) | 60.8 (1.67) | 71.3 (2.31) | 56.6 (1.42) |
| Heterogenous | 20 | 25.7 (0.44) | 42.3 (1.40) | 19.7 (0.32) | 23.6 (0.42) | 39.8 (1.01) | 17.3 (0.33) |
| | 25 | 29.9 (1.12) | 54.6 (0.97) | 19.8 (0.49) | 27.6 (0.95) | 53.5 (1.71) | 17.1 (0.33) |
| | 30 | 40.7 (2.35) | 77.4 (2.81) | 20.4 (0.31) | 35.0 (2.46) | 73.2 (2.69) | 17.2 (0.38) |

TABLE V
AVERAGE VEHICLE DELAYS AND THEIR STANDARD DEVIATIONS IN THE VARIOUS CONDITIONS

| Condition | SNR penalty | Vehicle delays (sec.) and their standard deviations | |
|---|---|---|---|
| | | Without correction | With correction |
| Baseline | | 40.58 (1.87) | |
| Homogenous | 0 | 41.27 (2.86) | 40.46 (2.58) |
| | 20 | 40.00 (2.22) | 40.48 (1.62) |
| | 25 | 41.54 (2.69) | 40.67 (1.75) |
| | 30 | 49.52 (4.35)* | 40.91 (1.70) |
| Heterogenous | 20 | 41.53 (2.34) | 40.73 (2.43) |
| | 25 | 42.16 (3.01) | 41.89 (2.64) |
| | 30 | 48.98 (5.13)* | 43.64 (2.76)* |

* p-value < 0.05 for difference from Baseline

section, to account for the communication distortions. When these corrections are applied, the additional delays are reduced substantially. In the homogeneous disruption conditions, the delays are similar to those in the Baseline scenario with all the SNR penalty levels including the 0 dB SNR penalty, for which the delay is only 1.0% higher than the Baseline. In the heterogenous disruption conditions, with the 30 dB SNR penalty, the delays are still significantly larger than in the Baseline, but only by 7.5%.

To better understand how the communication distortions affect the signal control algorithm and its performance, further analysis of the worst-case scenarios with 30 dB SNR penalties was conducted. It was already shown in Table V that the overall delays increase in these scenarios. The effects on the various signal groups are presented in Figure 4. The figure shows the percentage changes in the average delays, compared to the Baseline condition, for the various signal groups in four conditions: with the heterogenous or homogenous communication conditions, with and without applying the correction for communication distortions. The results demonstrate that the change in delay is not uniform. Without the correction for the distortions, in the homogenous communication condition, the increase in delays is largest for the signal groups from the South and West, while those from the East and North experience reduced delays. In the heterogenous communications condition, the West movements, which suffer the largest communication losses experience large delay increases, much larger than the traffic on the other approaches. Thus, the communication losses not only increase the overall delays, but they also change a distribution of the delays among the various signal groups. As the control logic reflects the priority regime among the various approaches to the intersection, this distortion has a negative impact on the traffic flow. In both environments, the application of the correction mechanism reduces the overall excess delays as well as the inequality of their distribution among the SGs. Most notably, in the heterogeneous condition, the additional delay to the West TR signal groups reduces from 92.4% to 35.2%.

Message losses may affect the control algorithm due to three fundamental phenomena:
1. The communication distortions affect the decision process taking place at the switch points between stages. At this point, the current stage is terminated and the algorithm selects the next one to be activated, which would be the stage with the highest score $w_{st}$ (see Eq. 3). The stage with the largest weighted demand may not be selected due to underestimation of its score. Thus, a provision of green time to the signal groups within this stage is postponed and the delays to the relevant vehicles increase.
2. The calculations of proportional scores $w_s^e$ (see Eq. 4) may also be affected. When these are underestimated, shorter maximum green durations would be set in the next cycle. Thus, shorter green times may be available to congested stages. When the proportional scores are overestimated, longer maximum green times may be allowed for stages that would not utilize them.
3. Loss of communication with the approaching vehicles may also result in a termination of the current stage when a demand for it is still sufficiently high to extend it.

Figure 5 depicts the extent that these events occur in the scenarios with 30 dB SNR penalties compared to the perfect communication scenario. They are presented in parallel to the changes in delay (Figure 5a), and all the figures reveal the

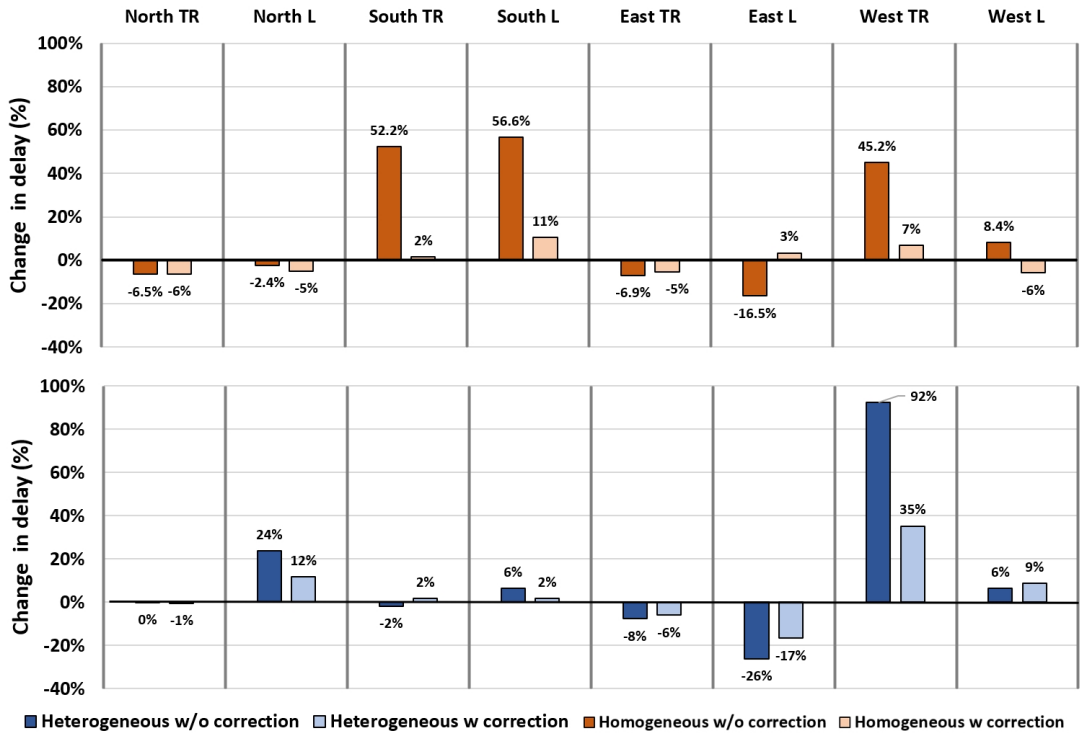
Fig. 4. Percentage change in the average delays for signal groups in the scenarios with 30 dB SNR penalties.

association between each type of event and its impact on the delays.

Figure 5b shows the frequencies of situations described in the first phenomenon above, where the next stage selected by the control algorithm is not the one that would have been selected under the perfect communication condition. As a result, some signal groups receive green time later in the cycle than they should have, while others receive it earlier and the difference between the frequencies of both types of events is presented. Although the order of the SG in the cycle is only one of the three main phenomena having an impact on the delay, it is expected that the greater the difference between the frequencies of late and early events, the greater its impact on the delay, and the correlation between the two ($R^2 = 0.68$) indicates this trend. As expected, the West signal groups tend to receive late green more frequently in the heterogenous communications condition. However, there is no clear pattern on which SGs gain or lose in the homogenous communication conditions. In both cases, the communication distortions correction reduces these events by about 40% and improves the equality of their distribution among the SGs.

The effect of errors in the allocation of maximum green times to stages on the actual green times, i.e. the second phenomenon above, is evaluated through calculation of the resulting green time loss to each SG (Figure 5c). A stage loses green time, compared to the perfect communication condition, when all the following conditions are met: the maximum green time allocated to it is shorter than it should have been, the maximum green time is reached, and there is an additional demand that would merit an extension of this stage. In contrast, a stage gains green time when the actual green time for the stage exceeds the maximum green time that should have been allocated to it. Figure 5c depicts the green time losses and gains in parallel to the delay. In many (although not all the cases) of the SGs and scenarios, the effect of time loss on the delay is observable ($R^2 = 0.65$). Losses in the homogeneous communication conditions are small, however, in the heterogeneous communication conditions, the West SGs that suffer higher communication distortions lose substantial green times, while the most other SGs gain green times. In both conditions, the introduction of the communication distortions correction substantially reduces the losses and gains: by 75% for the homogenous conditions and 83% for the heterogeneous conditions.

Finally, Figure 5d shows the numbers of vehicles that are delayed an additional cycle because their stage is not extended when it should have been as a result of communication distortions. As with the other effects of message losses, the West approach is negatively affected in the heterogeneous communication conditions, while there is no clear pattern in the homogeneous communication conditions. Naturally, this type of event is strongly correlated with the delays in the various SGs ($R^2 = 0.82$). The correction of communication distortions reduces the number of vehicles that are delayed a cycle due to communication losses by 67% and 68% for the homogeneous and heterogeneous conditions, respectively. These delayed vehicles are also more evenly distributed among the stage groups.

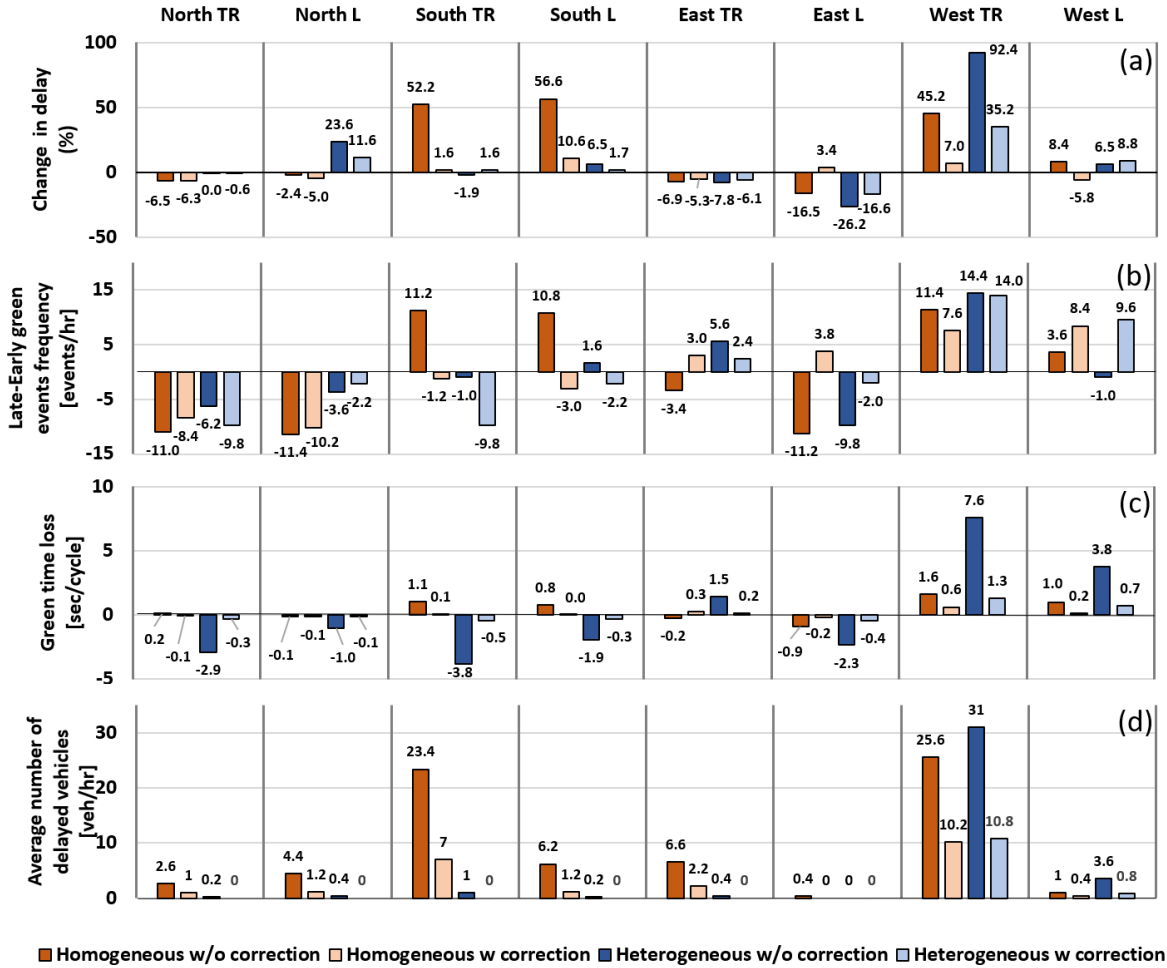

Fig. 5. (a) Change in delay; (b) Frequency of Late green time provision events minus frequency of Early green time provision; (c) Green time loss; (d) Number of delayed vehicles. (a) - (d) refer to scenarios with 30 dB SNR penalties compared to perfect communication scenarios.

VI. CONCLUSIONS

In this paper, the effect of communication distortions on the performance of a generic signalised intersection control algorithm in the connected vehicle environment was investigated. In this context, communication distortions have been previously ignored in the literature. A model system that integrated traffic flow model with a communication model was developed and applied to an intersection control scenario. The results show that considering the standard frequency of sending CAM messages of 1s in the DSRC-VANET communication network, the control algorithm is relatively robust to lost messages, even if the fraction of the lost messages is more than doubled compared to the normal conditions. However, if the fraction of the lost messages exceeds a critical level, reaching 75%, the vehicle delays start to grow dramatically. This behaviour is largely independent of whether all the intersection approaches or only one intersection approach experience communications distortions. Even if all the intersection approaches experience the similar message loss rates, the effects do not fully cancel out. Asymmetries in average flows at the intersection approaches and the competition of the traffic flows for the common intersection area are translated into uneven increases in vehicle delays. Hence, communication distortions resulted in more unfair distribution of vehicle delays among the signal groups. Using previously received data and simple assumptions, it was shown that corrections for the lost information can be made. When the communication distortions affected a single intersection approach, the intersection control deploying this simple correction showed a significant decrease in the average vehicle delays. Even better improvements were observed with the communications fault correction when the communication distortions impacted all the intersection approaches. The stochastic nature of traffic flow patterns poses a challenge when attempting to explain every single situation occurring in the intersection and its impact on the overall performance. Nevertheless, a comparative analysis of the control actions taken under imperfect information to the ones that should have been taken reveal a correlation between the incorrect actions and the delay of the various SGs.

The simulation results illustrate that a traffic control algorithm, which can identify the lost messages and makes the short-term predictions of the vehicle movements, partially compensates for temporal communication distortions and reduces their impact on vehicle delays. This allows to decrease either the required reliability level of the communication system or the volume of the vehicle-to-infrastructure communication, or both.

Future research should identify suitable methods, such as machine learning, and potentially useful data sources to be used to improve the reconstruction of the distorted communication. A higher applicability of the results can be achieved by improving the understanding of what is the minimum required information (i.e. the frequency of the sent CAM messages and the acceptable MLR) to make sure that a sufficient performance of the intersection control can be maintained and how it depends on the traffic (e.g. CV penetration level, time-varying flow intensities) and intersection configurations (e.g. consideration of other transportation modes, different saturation levels of the intersection). The results may also be extended to networks of coordinated traffic signals to allow continuous traffic flow over several intersections. Moreover, the question of how to handle the potential problems with the communication reliability in the context of the C-V2X networks such as LTE or 5G should be investigated.

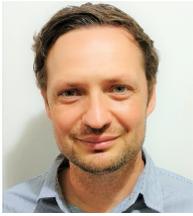
**Ilya Finkelberg** received the M.Sc. degree from the Technion – Israel Institute of Technology in 2012. He is a researcher and a traffic engineer in the Transportation Research Institute since 2008. He specializes in traffic management and urban mobility applied research. He is a member of the development team of AVIVIM – traffic management system of Tel Aviv and Haifa municipalities.

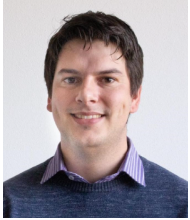
**Tibor Petrov** received the PhD. degree in Telecommunications from the University of Žilina (2018). He is a Researcher in Intelligent Transport Systems with the University of Žilina, Department of International Research Projects – ERAdiate+. His research activities include vehicular and cellular communication networks, vehicular network applications and computer modelling of cooperative Intelligent Transportation Systems.

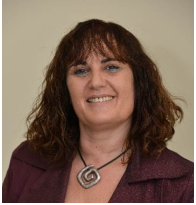
**Ayelet Gal-Tzur** holds a B.Sc and M.Sc. degrees in Industrial and Management Engineering and a D.Sc degree in traffic management, all from the Technion, Israel. She is leading the traffic mobility research team at the Transportation Research Institute of the Technion and she is also a senior lecturer at the Ruppin Academic Center. Her research focuses on sustainable mobility and urban traffic management with particular interest in ICT-based information sources, decision support methodologies and big data analytics in transportation.

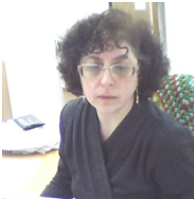
**Nina Zarkhin** received B.S. and M.A. degrees in applied mathematics from the Technical University of Kazan, Russia (faculty of computer science). She has been a member of the urban mobility research team at the Transportation Research Institute in the Technion since 2007. She took part in various traffic mobility research projects, including CONDUITS project in collaboration with several European municipalities. Her focus is on the software development aspect of traffic mobility applications and methodologies.

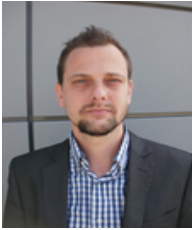
**Peter Počta** received his M.S. and Ph.D. degrees from University of Žilina, Faculty of Electrical Engineering, Slovakia in 2004 and 2007, respectively. He is currently a Full Professor at the Department of Multimedia and Information-Communication Technologies of the University of Žilina and is involved with International Standardization through the ETSI TC STQ as well as ITU-T SG12. His research interests include speech, audio, video and audiovisual quality assessment, speech intelligibility, multimedia communication and QoE management.

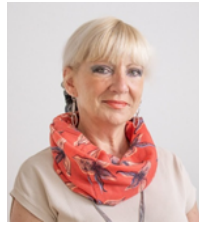
**Tatiana Kováčiková** has been Head of ERAdiate+ at the University of Zilina, Slovakia, since July 1, 2019. From October 1, 2017 till June 30, 2019 she was the ERAchair Holder for Intelligent Transport Systems (ITS) at the University of Zilina. In 2016, she was nominated a National delegate for H2020 PC on Smart, Green and Integrated Transport. She has been active in ICT&ITS standardisation for more than 15 years, currently she represents Slovakia in CEN TC 278 on ITS. From June 2013 till October 2015, she held the position of the Head of Science Operations at the COST Association in Brussels. In 2013, she was appointed Full Professor in Applied Informatics. Her research interests include ICT and Intelligent Transport Systems (ITS), in particular network architectures, services and applications..

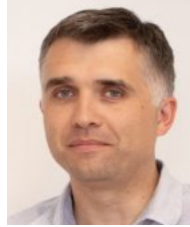
**Ľuboš Buzna** received the Ph.D. degree from the University of Žilina, in 2003. In the past, he worked as a Postdoctoral Researcher at several institutions, including the University of Barcelona, ETH Zurich, and TU Dresden. He is currently a Professor of Applied Informatics with the University of Žilina. His research is focused on the development of the optimization algorithms applied to transportation and other complex systems.

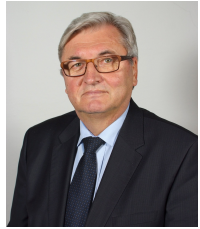
**Milan Dado** is Full Professor with the Department of Multimedia and ICT at the University of Zilina. He has been actively involved in European research and education programs (TEMPUS, COST, LEONARDO, Socrates, 5th, 6th and 7th Framework Program, H2020, European University association projects…) and has managed national projects related to information and communication technologies, intelligent transportation systems, regional innovation strategies and e-learning. Main milestones for his international activities were stays abroad e.g. Two-month stay at the York University Canada, Northern Telecom and Bell Canada in 1993, Six-month stay at the Royal Institute of Technology Stockholm in 1990 and Six -month stay at the Vienna University of Technology in 1981-1982. He has visited many other foreign institutions during last 30 years.

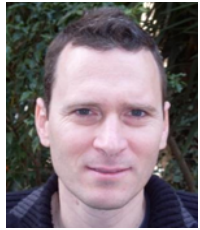
**Tomer Toledo** is an Associate Professor at the Faculty of Civil and Environmental Engineering and the Head of the Transportation Research Institute at the Technion – Israel Institute of Technology. He holds B.Sc. and M.Sc. degrees in Civil Engineering from the Technion and a Ph.D. in Transportation Systems from MIT. His research focuses on driver behaviour, traffic modelling and simulation, intelligent transportation systems and transportation network analysis.